# Massive MIMO 5G Cellular Networks: mm-wave vs. μ-wave Frequencies

Stefano Buzzi[1], *and* Carmen D'Andrea

University of Cassino and Lazio Meridionale, Italy

{buzzi, carmen.dandrea}@unicas.it

## Abstract

Enhanced mobile broadband (eMBB) is one of the key use-cases for the development of the new standard 5G New Radio for the next generation of mobile wireless networks. Large-scale antenna arrays, a.k.a. Massive MIMO, the usage of carrier frequencies in the range 10-100 GHz, the so-called millimeter wave (mm-wave) band, and the network densification with the introduction of small-sized cells are the three technologies that will permit implementing eMBB services and realizing the Gbit/s mobile wireless experience. This paper is focused on the massive MIMO technology; initially conceived for conventional cellular frequencies in the sub-6 GHz range (μ-wave), the massive MIMO concept has been then progressively extended to the case in which mm-wave frequencies are used. However, due to different propagation mechanisms in urban scenarios, the resulting MIMO channel models at μ-wave and mm-wave are radically different. Six key basic differences are pinpointed in this paper, along with the implications that they have on the architecture and algorithms of the communication transceivers and on the attainable performance in terms of reliability and multiplexing capabilities.

## 1. Introduction

Fifth-generation (5G) wireless network are expected to provide 1000x improvement on the supported data-rate, as compared to current LTE networks. Such an improvement will be mainly achieved through the concurrent use of three factors [1]: (a) the reduction in the size of the radio-cells, so that a larger data-rate density can be achieved; (b) the use of large-scale antenna arrays at the Base Stations (BSs), i.e., massive MIMO [2], so that several users can be multiplexed in the same time-frequency resource slot through multiuser MIMO (MU-MIMO) techniques; and (c) the use of carrier frequencies in the range 10-100 GHz, a.k.a. millimeter-waves (mm-waves) [3], so that larger bandwidths become available. The factor (a), i.e. the densification of the network, is actually a trend that we have been observing for some decades, in the sense that the size of the radio-cells has been progressively reduced over time from one generation of cellular networks to the next one. Differently, factor (b) can be seen as a sort of 4.5G technology, in the sense that latest 3GPP LTE releases already include the possibility to equip BS with antenna arrays of up to 64 elements. This trend will certainly continue in the future 5G New Radio standard, since the potentialities of massive MIMO are currently being tested worldwide in a number of real-world experiments (see, for instance, [4] and [5]). The use of mm-waves, on the contrary, is a more recent technology – at least as far as wireless cellular applications are concerned, and, although there is no doubt that future cellular networks will rely on them, it can be certainly classified as a true 5G technology.

---

[1] Corresponding Author: Stefano Buzzi, University of Cassino and Lazio Meridionale, Dept. Of Electrical and Information Engineering, Via G. Di Biasio, 43, I-03043 Cassino (FR), Italy. E-mail: buzzi@unicas.it .

Focusing on the massive MIMO technology, most of the research and experimental work has mainly considered its use at conventional (i.e. sub-6 GHz) cellular frequencies. We denote here such a range of frequencies as μ-wave, to contrast them with the above-6 GHz frequencies that we will denote as mm-wave[2]. Only recently, the combination of the massive MIMO concept with the use of mm-wave frequency bands has started being considered [6], [7]. As a matter of fact, the channel propagation mechanisms at μ-wave frequencies are completely different from those at mm-waves. As an instance, at μ-wave in urban environments the so-called rich-scattering environment is observed [8], thus implying that the MIMO channel is customarily modeled as the product of a scalar constant, taking into account the shadowing effects and the path-loss, times a matrix with i.i.d. (independent and identically distributed) entries. At mm-waves, instead, propagation is mainly based on Line-of-Sight (LOS) propagation and on one-hop reflections, and blockage phenomena are more frequent; to capture these mechanisms, a finite-rank clustered channel model is thus usually employed [9-11]. This paper compares massive MIMO systems at μ-wave with massive MIMO systems at mm-wave. We observe that these two different channel models have key implications on the achievable performance, on the multiplexing capabilities of the channels themselves, on the beamforming strategies that can be employed, on the transceiver algorithms and on the adopted channel estimation procedures. Six key differences between massive MIMO systems at μ-wave and massive MIMO systems at mm-wave are thus identified and critically discussed.

The rest of this paper is organized as follows. Section 2 contains the description of the considered transceiver model and of the massive MIMO channel models at μ-waves and at mm-wave frequencies. Section 3, the *core* of the paper, is divided in six subsections, each one describing a key difference between the massive MIMO channels at μ-wave and at mm-wave frequencies; numerical results are also shown here in order to provide experimental evidence of the theoretical discussion. Finally, concluding remarks are given in Section 4.

## 2. System and channel models

In this Section, we briefly illustrate the considered transceiver architecture and review the main characteristics of the MIMO wireless channel at μ-wave and mm-wave carrier frequencies.

We consider a MIMO wireless link with $N_T$ antennas at the transmitter and $N_R$ antennas at the receiver. We denote by $d$ the distance between transmitter and receiver, and by $M$ the number of transmitted parallel data streams (i.e., the multiplexing order). The considered transceiver model is reported in Fig. 1.

### 2.1 μ-wave channel model

Assuming frequency-flat fading (i.e. either multipath may be neglected or it has been nulled through the use of OFDM modulation), at channel frequencies below 6 GHz, the propagation channel is customarily modelled through an $(N_R \times N_T)$-dimensional matrix, whose $(i,j)^{th}$ entry, $[\boldsymbol{H}_\mu]_{i,j}$ has the following structure [12], [13]:

$$[\boldsymbol{H}_\mu]_{i,j} = \sqrt{\beta} g_{i,j}, \qquad (1)$$

---

[2] Notice however that, strictly speaking, the mm-wave bands correspond to carrier frequencies larger than 30 GHz.

where $g_{i,j}$ represents the small-scale (fast) fading between the $i^{th}$ receive antenna and the $j^{th}$ transmit antenna, and $\beta$ represents the (slow) large-scale fading (shadowing) and the path-loss between the transmitter and the receiver. In a *rich scattering environment,* the coefficients $g_{i,j}, i = 1, \ldots, N_R, j = 1, \ldots, N_T$ are i.i.d. $\mathcal{CN}(0,1)$ random variables. The factor $\beta$ is assumed constant across the transmit and receive antennas (i.e., it does not depend on the indices $i, j$), and is usually expressed as:

$$\beta = PL\, 10^{0.1\sigma_{sh}z}, \qquad (2)$$

where $PL$ represent the path loss and $10^{0.1\sigma_{sh}z}$ represents the shadow fading with the standard deviation $\sigma_{sh}$ and $z \sim \mathcal{N}(0,1)$. With regard to the path loss $PL$, several models have been derived over the years, based on theoretical models and/or on empirical heuristics. According to the popular three-slope model [13], [14], the path loss in logarithmic units is given by:

$$PL = \begin{cases} -L - 35 \log_{10} d, & \text{if } d > d_1 \\ -L - 15 \log_{10} d_1 - 20 \log_{10} d, & \text{if } d_0 < d \leq d_1 \\ -L - 15 \log_{10} d_1 - 20 \log_{10} d_0, & \text{if } d \leq d_0 \end{cases} \qquad (3)$$

where

$$L = 46.3 + 33.9 \log_{10} f - 13.82 \log_{10} h_T - (1.1 \log_{10} f - 0.7) h_R + 1.56 \log_{10} f - 0.8, \qquad (4)$$

with $f$ the carrier frequency in MHz, $h_T$ the transmitter antenna height in meters, and $h_R$ the receiver antenna height in meters. Given the fact that the small-scale fading contribution to the entries of the matrix $\mathbf{H}_\mu$ are i.i.d random variates, the channel matrix has full-rank with probability 1, and its rank is equal to the minimum value between $N_T$ and $N_R$.

## 2.2 mm-wave channel model

At mm-wave, propagation mechanisms are different from those at μ-wave. Indeed, path-loss is much larger, while diffraction effects are practically negligible, thus implying that the typical range in cellular environments is usually not larger than 100 m, and the non Line-of-Sight component is mainly based on reflections. Moreover, signal blockages, due to the presence of macroscopic obstacles between the transmitter and the receiver, are much more frequent than at μ-wave frequencies. In order to catch these peculiarities, general consensus has been reached on the so-called clustered channel model [7], [15]–[18]. This model is based on the assumption that the propagation environment is made of $N_{cl}$ scattering clusters, each of which contributes with $N_{ray}$ propagation paths, plus a possibly present LOS component. Apart from the LOS component, the transmitter and the receiver are linked through single reflections on the $N_{cl}$ scattering clusters. Assuming again frequency-flat fading, and focusing on a bi-dimensional model for the sake of simplicity, the baseband equivalent of the propagation channel is now represented by an ($N_R \times N_T$)-dimensional matrix expressed as:

$$\mathbf{H} = \gamma \sum_{i=1}^{N_{cl}} \sum_{l=1}^{N_{ray}} \alpha_{i,l} \sqrt{L(r_{i,l})}\, \mathbf{a}_r(\phi_{i,l}^r) \mathbf{a}_t^H(\phi_{i,l}^t) + \mathbf{H}_{LOS} \qquad (5)$$

In the above equation, we denote by $\phi_{i,l}^r$ and $\phi_{i,l}^t$ the angles of arrival and departure of the $l^{th}$ ray in the $i^{th}$ scattering cluster, respectively. The quantities $\alpha_{i,l}$ and $L(r_{i,l})$ are the complex path gain and the attenuation associated to the $(i,l)^{th}$ propagation path. Following [10], the attenuation $L(r_{i,l})$ of the $(i,l)^{th}$ path is written in logarithmic units as:

$$L(r_{i,l}) = -20 \log_{10}\left(\frac{4\pi}{\lambda}\right) - 10n \left[1 - b + \frac{bc}{\lambda f_0}\right] \log_{10}(r_{i,l}) - X_\sigma, \qquad (6)$$

with $\lambda$ the wavelength, $c$ the speed of light, $n$ the path loss exponent, $X_\sigma$ the zero-mean, $\sigma^2$ –variance Gaussian-distributed shadow fading term in logarithmic units, $b$ a system parameter, and $f_0$ a fixed reference frequency, the centroid of all the frequencies represented by the path loss model. The values for all these parameters for the four different use-case scenarios discussed in [10] (Urban Microcellular (UMi) Open-Square, UMi Street-Canyon, Indoor Hotspot (InH) Office, and InH Shopping Mall) are reported in Table 1. The complex gain $\alpha_{i,l} \sim \mathcal{CN}(0, \sigma_{\alpha_i}^2)$, with $\sigma_{\alpha_i}^2 = 1$ [15]. The factors $\boldsymbol{a}_r(\phi_{i,l}^r)$ and $\boldsymbol{a}_t(\phi_{i,l}^t)$ represent the normalized receive and transmit array response vectors evaluated at the corresponding angles of arrival and departure; for an uniform linear array (ULA) with half-wavelength inter-element spacing we have $\boldsymbol{a}_t(\phi_{i,l}^t) = \frac{1}{\sqrt{N_T}}\left[1, e^{-j\pi \sin \phi_{i,l}^t}, \ldots, e^{-j\pi(N_T-1)\sin \phi_{i,l}^t}\right]^T$. A similar expression can be also given for $\boldsymbol{a}_t(\phi_{i,l}^t)$. Finally, $\gamma = \sqrt{\frac{N_T N_R}{N_{cl} N_{ray}}}$ is a normalization factor ensuring that the received signal power scales linearly with the product $N_T N_R$. Regarding the LOS component, denoting by $\phi_{LOS}^r$ and $\phi_{LOS}^t$, the arrival and departure angles corresponding to the LOS link, we assume that

$$\boldsymbol{H}_{LOS} = I_{LOS}(d)\sqrt{N_T N_R L(d)} e^{j\vartheta} \boldsymbol{a}_r(\phi_{LOS}^r) \boldsymbol{a}_t^H(\phi_{LOS}^t). \qquad (7)$$

In the above equation, $\theta \sim U(0, 2\pi)$ and $I_{LOS}(d)$ is a random variate indicating the existence of a LOS link between transmitter and receiver. A detailed description of all the parameters needed for the generation of sample realizations for the channel model in Eq. (5) is reported in [9]. Comparing the channel model reported in Eq. (5) for mm-wave frequencies with the one reported in Eq. (1) for μ-wave frequencies, it is immediately evident that the channel in (5) is a parametric channel model whose rank is tied to the number of clusters and reflectors contributing to the transmitter-receiver link. Next Section provides an accurate description of the implications that these two radically different channel models have on the architecture and on the attainable performance of massive MIMO multiuser wireless systems operating at μ-wave and at mm-wave frequencies.

Table 1: Parameters for path loss model at mm-wave for four different use-case scenarios

| Scenario | Model Parameters |
|---|---|
| UMi Street Canyon LOS | $n = 1.98$, $\sigma = 3.1$ dB, $b = 0$ |
| UMi Street Canyon NLOS | $n = 3.19$, $\sigma = 8.2$ dB, $b = 0$ |
| UMi Open Square LOS | $n = 1.85$, $\sigma = 4.2$ dB, $b = 0$ |
| UMi Open Square NLOS | $n = 2.89$, $\sigma = 7.1$ dB, $b = 0$ |
| InH Indoor Office LOS | $n = 1.73$, $\sigma = 3.02$ dB, $b = 0$ |
| InH Indoor Office NLOS | $n = 3.19$, $\sigma = 8.29$ dB $b = 0.06$, $f_0 = 24.2$ GHz |
| InH Shopping Mall LOS | $n = 1.73$, $\sigma = 2.01$ dB, $b = 0$ |
| InH Shopping Mall NLOS | $n = 2.59$, $\sigma = 7.40$ dB $b = 0.01$, $f_0 = 39.5$ GHz |

## 3   mm-wave versus μ-wave massive MIMO

In the following, we highlight and discuss six key differences between μ-wave and mm-wave massive MIMO systems.

## Difference #1: mm-waves may be doubly massive

The idea of a large scale antenna array was originally launched by Marzetta in his pioneering paper [12] with reference to BSs. The paper showed that in the limit of large number of base station antennas small-scale fading effects vanish by virtue of channel hardening, and that channel vectors from the BS to the users tend to become orthogonal; consequently, plain channel-matched beamforming at the BS permits serving several users on the same time-frequency resource slot with (ideally) no interference, and the only left impairment is due to imperfect channel estimates due to the fact that orthogonal pilots are limited and they must be re-used throughout the network (this is the so-called pilot contamination effect, discussed in the following). Reference [12] considered a system where mobile users were equipped with just one antenna. Successive studies have extended the massive MIMO idea at μ-wave frequencies to the case in which the mobile devices have multiple antennas, but this number is obviously limited to few units. Indeed, at μ-wave frequencies the wavelength is in the order of several centimeters, and it is thus difficult to pack many antennas on small-sized user devices. At μ-waves, thus, massive MIMO just refers to BSs. Things are instead different at mm-waves, wherein multiple antennas are necessary first and foremost to compensate for the increased path-loss with respect to conventional sub-6 GHz frequencies. At mm-wave, the wavelength is on the order of millimeters, and, at least in principle, a large number of antennas can be mounted not only on the BS, but also on the user device. As an example, at a carrier frequency of 30 GHz the wavelength is 1 cm, and for a planar antenna array with λ/2 spacing, more than 180 antennas can be placed in an area as large as a standard credit card (8.5 cm x 5.5 cm); this number climbs up to 1300 at a carrier frequency of 80 GHz. This consideration leads to the concept of *doubly massive MIMO* system [7], which is defined as a wireless communication system where the number of antennas grows large at both the transmitter and the receiver. Of course, there are a number of serious practical constraints – e.g., large power consumption, low efficiency of power amplifiers, hardware complexity, ADC and beamformer implementation – that currently prevent the feasibility of a user terminal equipped with hundreds of antennas. Mobile devices with a massive number of antennas thus will not be available in few years, but, given the intense pace of technological progress, sooner or later they will become reality. As far as long-term forward-looking theoretical research is concerned, we believe that doubly-massive MIMO systems at mm-waves will be a popular research topic for years to come.

## Difference #2: Analog (beam-steering) beamforming may be optimal

One problem with massive MIMO systems is the cost and the complexity of the needed hardware to efficiently exploit a so large number of antennas. If fully digital beamforming is to be made, as many RF chains are needed as the number of antennas; consequently, also energy consumption grows linearly with the number of antennas. In order to circumvent this problem, lower complexity architectures have been proposed, encompassing, for instance, 1-bit quantization of the antenna outputs [19] and hybrid analog/digital beamforming structures [11], [18], [20], wherein an RF beamforming matrix (whose entries operate as simple phase shifters) is cascaded to a reduced-size digital beamformer. The authors of the paper [21] has shown that if the number of RF chains is twice the multiplexing order, then the hybrid beamformer is capable of implementing any fully digital beamformer. Now, while at μ-waves the use of hybrid beamformer brings an unavoidable performance degradation, at mm-waves something different happens in the limiting regime of large number of antennas by virtue of the different propagation mechanisms. Indeed, note that the channel matrix in Eq. (5) can be compactly re-written as:

$$\boldsymbol{H} = \gamma \sum_{i=1}^{N} \alpha_i \boldsymbol{a}_r(\phi_i^r) \boldsymbol{a}_t^H(\phi_i^t), \qquad (8)$$

where we have lumped into the coefficients $\alpha_i$ the path-loss term, and we have grouped the two summations over the clusters and the rays in just one summation, with $N$ being the number of propagation paths that from the transmitter arrive to the receiver. Given the continuous random location of the scatterers, the set of arrival angles will be different with probability 1, i.e. there is a zero probability that two distinct scatterers will contribute to the channel with the same departure and arrival angles. Since, for large number of antennas, we have that $\boldsymbol{a}_x^H(\phi_p^x)\boldsymbol{a}_x(\phi_q^x) \to 0$, provided that $\phi_p^x \neq \phi_q^x$, with $x = \{r, t\}$, we can conclude that for large $N_T$, the vectors $\boldsymbol{a}_t(\phi_i^t)$ for all $i = 1, \ldots N$, converge to an orthogonal set, and, similarly, for large $N_R$, the vectors $\boldsymbol{a}_r(\phi_i^r)$ for all $i = 1, \ldots N$ converge to an orthogonal set as well. Accordingly, in the doubly massive MIMO regime, the array response vectors $\boldsymbol{a}_r(\cdot)$ and $\boldsymbol{a}_t(\cdot)$ become the left and right singular vectors of the channel matrix, i.e. the channel representation (8) coincides with the singular-value-decomposition of the channel matrix. Under this situation, we have that *purely analog (beam-steering) beamforming becomes optimal*. Otherwise stated, we have two main consequences. First, in a single-user link, the channel eigendirections associated to the largest eigenvalues are just the beam-steering vectors corresponding to the arrival and departure angles associated with the predominant scatterers. This suggests that pre-coding and post-coding beamforming simply require pointing a beam towards the predominant scatterer, at the transmitter and at the receiver, respectively. Second, in a multiuser environment, assuming that the links between the several users and the BS involves separate scatterers and so different sets of arrival and departure angles[3], beam-steering analog beamforming automatically results in no-cochannel interference (in the limiting regime of infinite number of antennas) since the beams pointed towards different users tend to become orthogonal. Fig. 2 provides some experimental evidence of the above statements. We have considered a single-user MIMO link at mm-waves; the carrier frequency is 73 GHz, the transmitting antenna height is 15 m, while the receiving antenna height is 1.65 m. All the parameters needed for the generation of the mm-wave channel matrix in Eq. (5) are the ones reported in [9] for the "open square model". Fig. 2 shows the system spectral efficiency, measured in bit/s/Hz, versus the received signal to noise ratio (SNR), comparing the performances of the Channel Matched (CM) fully digital beamforming and the Analog (AN) beam-steering beamforming. With CM beamforming the pre-coding and post-coding beamformers are the left and singular eigenvectors of the channel matrix in Eq. (5) associated to the $M$ largest eigenvalues, respectively; with AN beamforming, instead, the pre-coding and post-coding beamformers are simply the array responses corresponding to the departure and arrival angles associated to the $M$ dominant scatterers, respectively. From the figure it is seen that AN beamforming achieves practically the same performance as CM beamforming for multiplexing order $M = 1$, even in the case of not-so-large number of antennas, while there is a small gap for $M = 3$; this gap is supposed to get reduced as the number of antennas increases.

## Difference #3: The rank of the channel does not increase with $N_T$ and $N_R$

At µ-wave frequencies, the i.i.d. assumption for the small-scale fading component of the channel matrix $\boldsymbol{H}$, guarantees that with probability 1 the matrix has rank equal to $\min(N_T, N_R)$. Consequently, as long as the rich-scattering environment assumption holds and the number of degrees of freedom of the radiated and scattered field is sufficiently high [22], the matrix rank increases linearly with the number of antennas. At mm-wave frequencies, instead, the validity of the channel model in Eq. (5) directly implies that, including the LOS component, the channel has at most rank $N_{cl}N_{ray} + 1$, since it is expressed as the sum of $N_{cl}N_{ray} + 1$ rank-1 matrices. This rank is clearly independent of the number of transmit and receive antennas, so, mathematically, as long as $\min(N_T, N_R) > N_{cl}N_{ray} + 1$, increasing the number of antennas has no effect on the channel rank. That said, however, intuition also suggests that, for increasing number of antennas, the directive beams become narrower and narrower, and more scatterers can be resolved, thus implying that channel rank increases (even though probably not linearly) with the number of antennas. We highlight however that this is a conjecture that would need experimental validation.

---

[3] This is a quite reasonable assumption for sufficiently spaced mobile user locations.

The described different behavior of the channel rank with respect to the number of antennas has a profound impact on the multiplexing capabilities of the channel. Indeed, while, for µ-wave systems, the increase in the channel rank leads to an increase of the multiplexing capabilities of the channel, in mm-wave systems the multiplexing capabilities depend on the number of scatterers in the propagation environment, while the number of antennas just contributes to the increase of the received power, which indeed can be shown to increase proportionally to the product $N_T N_R$. Fig. 3 provides experimental evidence of such a different behavior. The figure shows the system spectral efficiency for a mm-wave and for a µ-wave wireless MIMO links, for two different values of the number of receive and transmit antennas, and for three different values of the multiplexing order $M$. The parameters of the mm-wave channel are the same as those in Fig. 2, while, regarding the µ-wave channel, a carrier frequency equal to 1.9 GHz has been considered, the standard deviation of the shadow fading $\sigma_{sh}$ has been taken equal to 8 dB, while the parameters of three-slope path loss model in Eq. (3) are $d_1 = 50$ m and $d_2 = 100$ m. From Fig. 3 it is clearly seen that the µ-Wave channel has larger multiplexing capabilities than the mm-wave channel; the gap between the two scenarios is mostly emphasized for the large values of $M$ and for $N_R \times N_T = 100 \times 1000$.

## Difference #4: Channel estimation is simpler

In µ-Wave massive MIMO systems channel estimation is a rather difficult and resource-consuming task, since it requires the separate estimation of each entry of the matrix $\boldsymbol{H}$; it thus follows that in a multiuser system with $K$ users equipped with $N_R$ antennas each[4], the number of parameters to be estimated is $K N_R N_T$. The attendant computational complexity needed to perform channel estimation is a growing function of the number of used antennas. Additionally, the increase of the number of antennas $N_R$ at the mobile devices has a direct impact on the network capacity. Indeed, let $\tau_c$ denote the duration (in discrete samples) of the channel coherence time and $\tau_p$ the length (again in discrete samples) of the pilot sequences used on the uplink for channel estimation; since the pilot sequences length must be a fraction (typically no more than ½) of the channel coherence length, and since the use of orthogonal pilots across users requires that $K N_R \leq \tau_p < \tau_c$ , it is readily seen that we have a physical bound on the maximum number of users and on the number of transceiver antennas at the mobile device. Such a bound is the main underlying motivation for the fact that a considerable share of the available literature on massive MIMO systems at µ-waves focuses on the case of single-antenna mobile devices, so that with $N_R = 1$ the number of users $K$ can be taken larger. Additionally, to increase the number of supported users, pseudo-orthogonal pilots with low cross-correlation are used, even though this leads to the well-known pilot contamination problem that, as discussed in the sequel, is the ultimate performance limit in µ-wave massive MIMO systems [12].

At mm-wave frequencies, instead, the clustered channel model of Eq. (5) is basically a parametric model, and the number of parameters is essentially independent of the number of antennas. Based on this consideration, the computational complexity of the channel estimation schemes at mm-waves may be smaller than that at µ-waves. Channel estimation for mm-wave frequencies is a research track that is currently under development, whereas for µ-wave is a rather mature area. Among the several existing approaches to perform channel estimation at mm-wave, the most considered ones rely either on compressed sensing or on subspace methods. As an example, reference [23] shows that at mm-waves, for increasing number of antennas, the most significant components of the received signal lie in a low-dimensional subspace due to the limited angular spread of the reflecting clusters. This low-dimensionality feature can be exploited in order to obtain channel estimation algorithms based on the sampling of only a small subset rather than of the whole number of antenna elements. Consequently, channel estimation can be performed using a reduced number (with respect to the number of receive antennas) of required RF chains and A/D converters at receiver front-end. Reference [23], instead, develops subspace-based channel estimation methods exploiting channel reciprocity in TDD systems, using the well-known Arnoldi iteration, and explicitly taking into account the adoption of hybrid analog/digital beamforming structures at the transmitter and at the receiver. Subspace methods are particularly attractive in those situations where it is of interest to estimate the principal left and right singular eigenvectors of the channel matrix $\boldsymbol{H}$, which, in the doubly massive MIMO regime, are well-approximated by the array response vectors corresponding to the dominant scatterers. Applying, as done in [25], fast subspace estimation algorithms such as the Oja's one [26], the dominant channel eigenvectors can be directly obtained by the sample estimate of the data covariance matrix, with no need to directly estimate the whole channel matrix $\boldsymbol{H}$.

---

[4] $N_T$ here denotes the number of antennas at the BS.

Figures 4 and 5 report numerical results concerning channel estimation at μ-wave and at mm-wave channel frequencies. In particular, both figures report the spectral efficiency versus the received SNR for two different antenna configurations and by contrasting the case of perfect channel state information (CSI) with the case in which the channel is estimated based on training pilots. In both figures a single-user MIMO link is considered, and channel estimation is carried out assuming that each transmit antenna sends an orthogonal pilot. The number of signaling intervals devoted to channel estimation coincides with the number of transmit antennas – note that this is the minimum possible duration in order to be able to send orthogonal pilots. Channel estimation at μ-wave frequencies (Fig. 4) is made using the linear minimum mean square errors criterion (see [27]), while at mm-wave frequencies (Fig. 5) the approximate maximum likelihood (AML) algorithm of [23] and the orthogonal Oja (OOJA) algorithm [25] are used. Comparing the figures, it is clearly seen that the gap between the case of estimated channel and the case of perfect CSI is smaller at mm-wave frequencies, especially when the OOJA algorithm is considered. Conversely, this gap is larger at μ-waves, and it grows with the dimension of the user antenna arrays. This behavior can be intuitively explained by virtue of the parametric form of the mm-wave channel model (see Eq. (5)), which permits the development of efficient channel estimation algorithms.

## Difference #5: Pilot contamination can be less critical

Pilot contamination is the ultimate disturbance in massive MIMO systems operating at μ-waves. As already discussed in the previous paragraph, the impossibility to have a number of orthogonal pilots larger than the number of signaling intervals devoted to channel estimation leads to the use of pseudo-orthogonal, low cross-correlation sequences. Accordingly, in a massive MIMO system, when, in the uplink training phase, the MSs transmit their own pilot sequences to enable channel estimation at the BSs, every BS learns not only the channel from the intended MS, but also small pieces of the channels from the other MSs using pilots that are correlated to the one used by the intended MS. This phenomenon, in turn, causes a saturation in the achieved Signal-to-Interference plus Noise-Ratio (SINR) both in the downlink and in the uplink. The deceitful nature of pilot contamination was unveiled by Marzetta in his landmark paper [12] and since then, many authors have deeply investigated its effects and proposed strategies to counterbalance its effects [28], [29], [30]. All of these papers deal with the case of a μ-wave massive MIMO system.

Pilot contamination at mm-wave frequencies is instead a much less-studied topic (some initial results are reported in [31]). This is in part due to the fact that massive MIMO at mm-wave is a more recent research topic than massive MIMO at µ-wave. On the other hand, it may be envisioned that pilot contamination may be less critical at mm-waves than it has revealed at µ-waves, mainly for the short-range nature of mm-wave links. In particular, while the range of µ-wave links can be in the order of thousands of meters, the range for mm-wave links will be more than one order of magnitude smaller, due to increased path-loss and to a larger relevance of signal blockages. Mm-wave frequencies will be used for short-range communications in small cells, which, by nature, usually serve a smaller number of users than conventional micro-cells and macro-cells. So, on one hand, the signals transmitted by the MSs during uplink training fade rapidly with the distance, and thus they should not be a serious impairment to surrounding BSs learning the channel from their intended MSs; on the other hand, the reduced number of users in each cell will lead to a less severe shortage of orthogonal pilots. The results in [31] seem to confirm such increased resilience of mm-waves to the pilot contamination problem.

## Difference #6: Antenna diversity/selection procedures are less effective

The i.i.d. nature of the fast fading component in the MIMO channel matrix at µ-waves in Eq. (1) leads to a monotonic increase, with the number of antennas, of the diversity order that can be attained. In particular, an $N_R \times N_T$ channel brings a diversity order equal to $N_R \times N_T$, thus implying that the average error probability decreases to a zero, in the limit of large Signal-to-Noise Ratio (SNR), as $SNR^{-N_R N_T}$. Such a diversity order can be attained through a simple antenna selection procedure by picking the transmit and receive antennas corresponding to the entry with the largest magnitude in the channel matrix $\boldsymbol{H}$. Looking at this fact from a different perspective, we can recall the well-known probability result stating that the maximum of a set of positive i.i.d. random variables taking value in the interval $[0, +\infty)$, becomes unbounded as the cardinality of the set diverges: as a consequence, for increasing number of antennas, the probability of observing a very large entry in the channel matrix rapidly increases. The open literature is rich of studies exploiting this peculiarity of µ-wave MIMO channels and proposing diversity techniques based on antenna selection procedures (see, e.g., [32], [33]).

At mm-waves, instead, given the parametric channel model of Eq. (5), a different behavior is observed. In particular, the entries of the matrix channel have no longer an i.i.d. component, and this implies that the maximum of the magnitudes of the entries of $\boldsymbol{H}$ grows at a much reduced pace. As a consequence, diversity techniques using antenna selection procedures are less effective.

As an experimental evidence of this fact, Fig. 6 provides the following parameter in Eq. (9), for different values of $N_R \times N_T$, and for both the µ-wave and the mm-wave channel models:

$$\eta = \frac{\max_{i,j} |\boldsymbol{H}_{i,j}|^2}{tr(\boldsymbol{H}^H \boldsymbol{H})/N_T N_R} \; . \qquad (9)$$

The quantity $\eta$ is the ratio between the largest squared magnitude among the entries of $\boldsymbol{H}$, and the average squared magnitude. Of course, the larger $\eta$, the more unbalanced are the magnitudes of the entries of the channel matrix, since $\eta$ basically measures how far is the largest entry in $\boldsymbol{H}$ from the average magnitude. Inspecting Fig.6, it is clearly seen that the parameter $\eta$ is in general an increasing function of the number of antenna elements, but it grows much more rapidly in the case of µ-wave channels.

# 4  Conclusions

This paper has outlined a critical comparison between massive MIMO systems at mm-waves and at µ-waves. Six key differences have been outlined, and their implications on the transceiver architecture and on the attainable performance have been discussed and validated also through the result of computer simulations. Among the discussed differences, we believe that the most disruptive one is the difference #1, i.e. the fact that MIMO systems may be doubly massive at mm-waves. Indeed, while it has been shown that the use of large-scale antenna arrays has not an as beneficial impact on the system multiplexing capabilities as it has at µ-wave frequencies, the availability of doubly massive MIMO wireless links will enable the generation of very narrow beams, resulting thus in reduced co-channel interference to other users using the same time-frequency resources. Another key advantage of doubly massive MIMO systems at mm-waves is the fact that the computational complexity of channel estimation weakly depends on the number of antennas, especially for the case in which analog (beam-steering) beamforming strategies are used. While already discussed, massive MIMO at µ-wave frequencies is gradually entering in 3GPP standards, mm-waves and in particular massive mm-wave MIMO systems are still under heavy investigation, both in academia and industry. It is however anticipated that sooner or later a technology readiness level will be reached such that they will be included in 3GPP standards. The authors of this paper hope that this article will help to move us forward along this road.

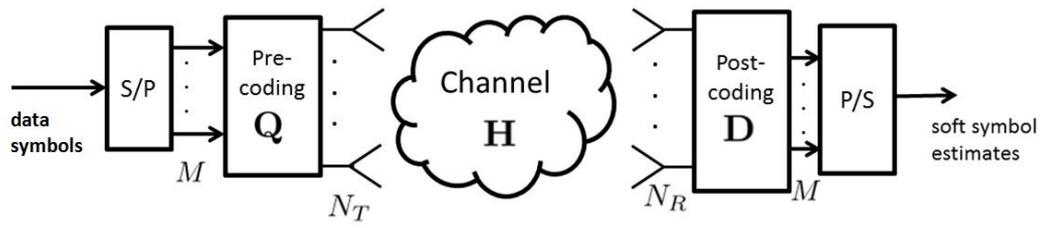

Fig. 1: The considered transceiver model

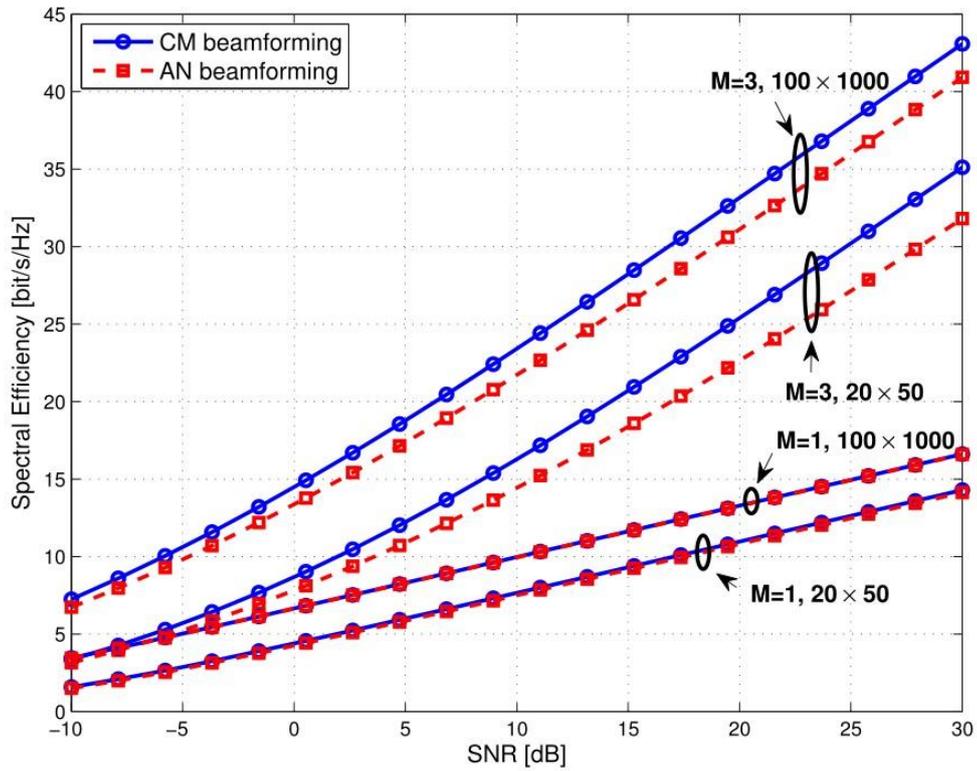

Fig. 2: Spectral Efficiency of a mm-wave MIMO wireless link versus received SNR for CM-FD beamforming and AN (beam-steering) beamforming, for two different values of the number of transmit and receive antennas and of the multiplexing order of the system.

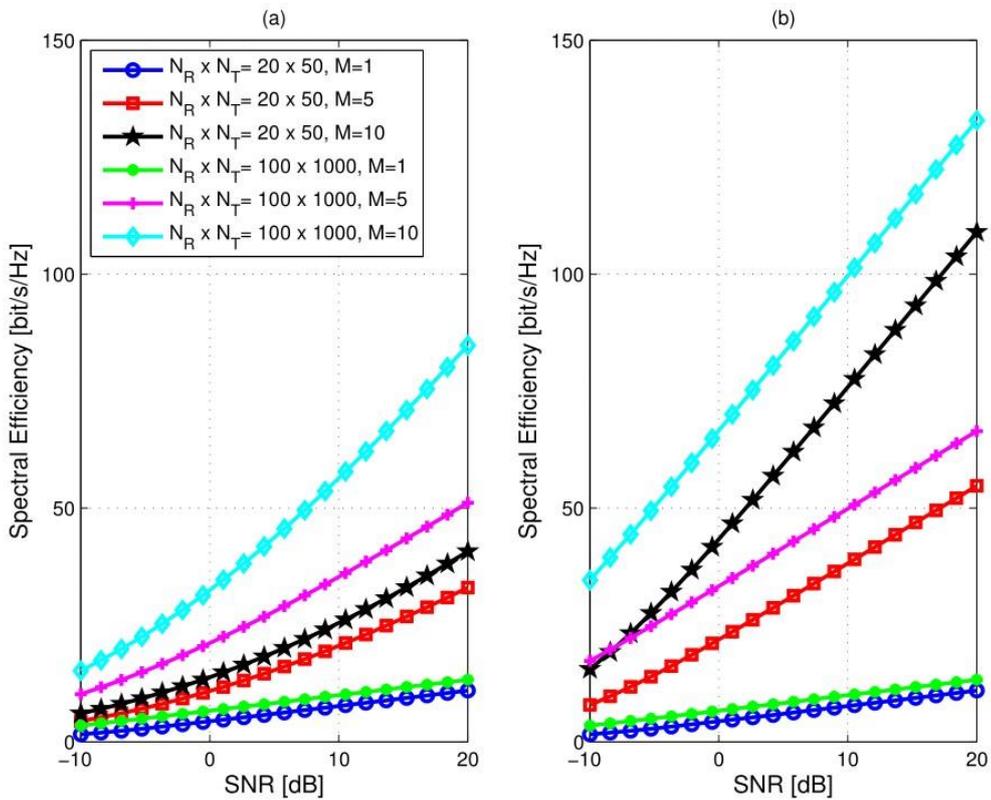

Fig. 3: (a) Spectral Efficiency versus received SNR for mm-wave channel varying the number of transmit and receive antennas and multiplexing order, (b) Spectral Efficiency versus received SNR for μ-wave channel varying the number of transmit and receive antennas and multiplexing order.

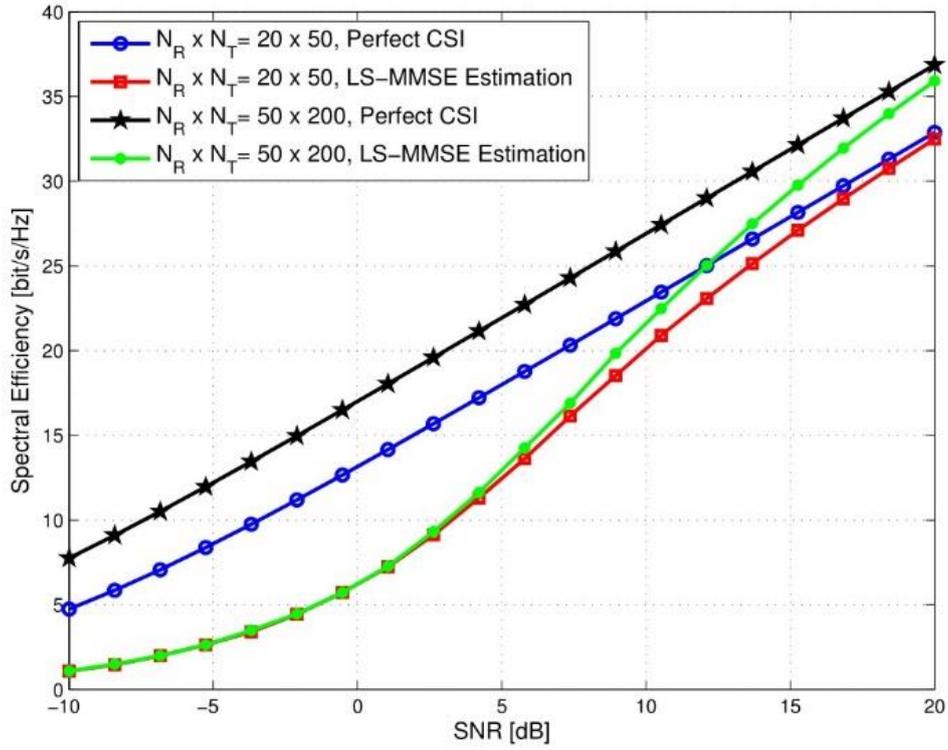

Fig. 4: Spectral Efficiency versus received SNR with Perfect CSI and Imperfect CSI, with LS-MMSE Algorithm for the Estimation of μ-wave Channel. The multiplexing order is 3.

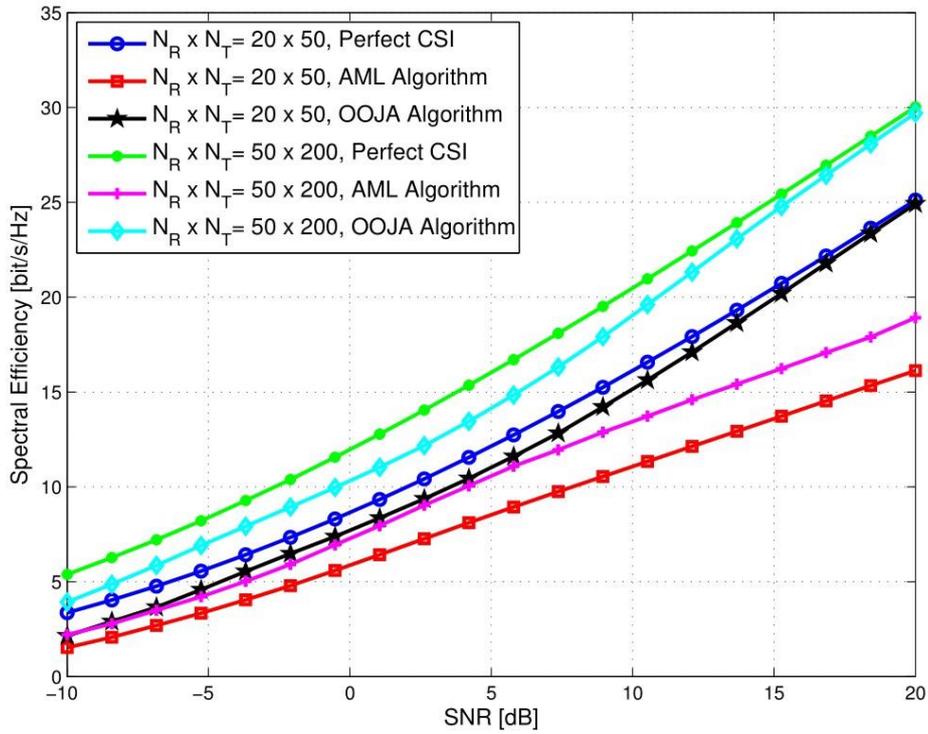

Fig. 5: Spectral Efficiency versus received SNR with Perfect CSI and Imperfect CSI, with AML Algorithm and OOJA Algorithm for the Estimation of mm-wave Channel. The multiplexing order is 3.

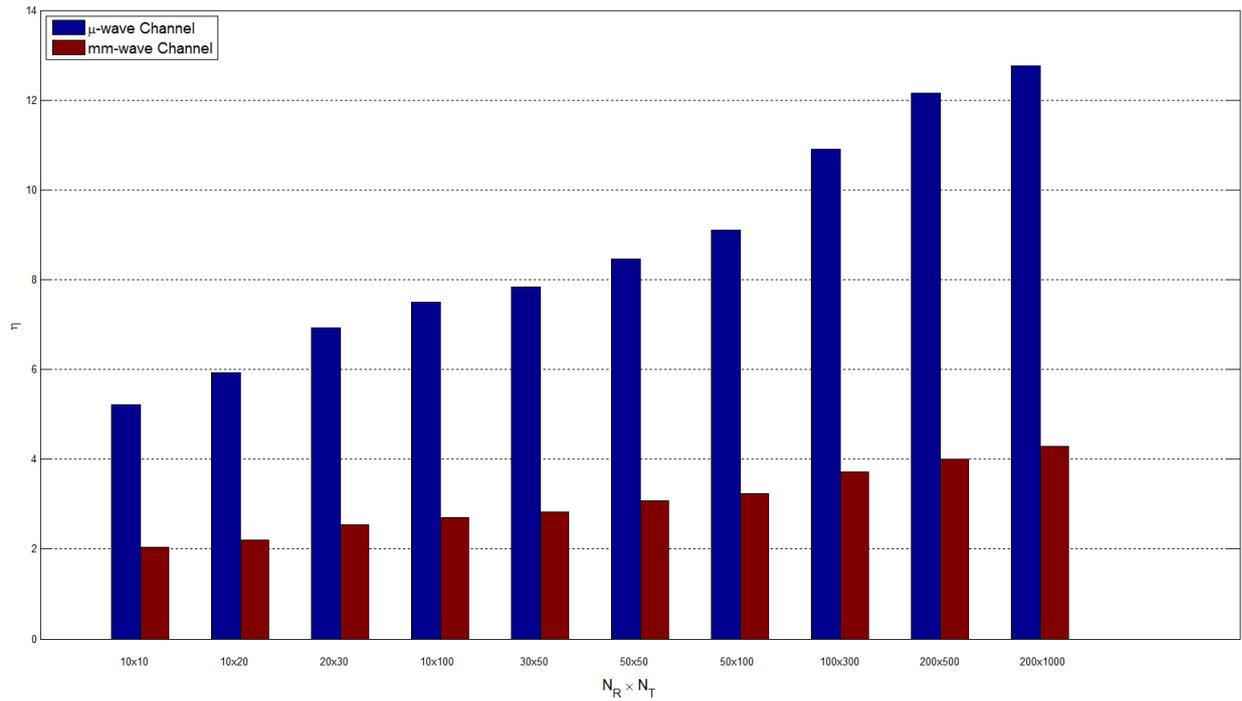

Fig. 6: Value of the performance measure η defined in Eq. (9) for several antenna array sizes, for the mm-wave and the μ-wave channel.

**STEFANO BUZZI** is currently an Associate Professor at the University of Cassino and Lazio Meridionale. He received the Ph.D. degree in Electrical and Computer Engineering from the University of Naples "Federico II" in 1999, and has had short-term research appointments at Princeton University, Princeton (NJ), USA in 1999, 2000, 2001 and 2006. He is a former Associate Editor of the IEEE Signal Processing Letters and of the IEEE Communications Letters, while is currently serving as an Editor for the IEEE Transactions on Wireless Communications. Dr. Buzzi's research interests are in the broad field of communications and signal processing, with emphasis on wireless communications. He has co-authored about 150 technical peer-reviewed journal and conference papers, and among these, the highly-cited survey paper "What will 5G be?" (IEEE JSAC, June 2014) on 5G wireless networks.

**CARMEN D'ANDREA** was born in Italy on 16 July 1991. She received the B.S. and M.S. degrees, both with honors, in Telecommunications Engineering from University of Cassino and Lazio Meridionale in 2013 and 2015, respectively. She is currently with the Department of Electrical and Information Engineering at the University of Cassino and Lazio Meridionale, pursuing the Ph.D. degree in Electrical and Information

Engineering. Her research interests are focused on wireless communication and signal processing and her current focus is on mm-wave communications and massive MIMO systems.